\documentclass[twocolumn]{aastex631}

\usepackage{pgfplots}
\usepackage{cancel}
\usepackage{ulem}
\usepackage{float}
\usepackage{times}
\usepackage{mathptmx}  

\date{May 16, 2026}

\begin{document}

\title{On the Nature of Candle-Flame-Shaped Solar Flares and Sub-Alfvénic Supra-Arcade Plasma Downflows}

\author[0000-0002-1876-1372]{Ivan Oparin}
\author[0009-0005-5227-5633]{Sabastian Fernandes}
\author[0000-0002-0660-3350]{Bin Chen}
\affil{Center For Solar-Terrestrial Research, New Jersey Institute of Technology, Newark, NJ 07102, USA}
\author[0000-0002-9258-4490]{Chengcai Shen}
\affiliation{Center for Astrophysics | Harvard \& Smithsonian, 60 Garden Street, Cambridge, MA, 02138, USA}
\author[0000-0001-5278-8029]{Xiaocan Li}
\affiliation{Los Alamos National Laboratory, Los Alamos, NM 87545, USA}
\author[0000-0003-4315-3755]{Fan Guo}
\affiliation{Los Alamos National Laboratory, Los Alamos, NM 87545, USA}
\author[0000-0003-2872-2614]{Sijie Yu}
\affil{Center For Solar-Terrestrial Research, New Jersey Institute of Technology, Newark, NJ 07102, USA}





\begin{abstract}

Certain solar flares exhibit a distinctive candle-flame or cusp-shaped feature above the bright flare arcade visible in extreme ultraviolet (EUV) and X-ray channels sensitive to high-temperature plasma. The presence of a cusp-like structure is generally regarded as a key piece of morphological evidence for magnetic reconnection to power explosive energy release in solar flares. In addition, downward-propagating plasma flows above the flare arcade have often been interpreted as outflows driven by magnetic reconnection. However, the relationship between the observed candle-flame-shaped morphology and the underlying magnetic field geometry for reconnection remains unclear. 
Likewise, the observed speed of the plasma downflows has been found to be too low compared to the upstream Alfv\'en speed predicted by reconnection theories. 
With the help of a recently developed three-dimensional magnetohydrodynamics model, we examine the locations where magnetic topology changes from antiparallel to closed (Y-points) in a candle-flame-shaped flare, compare the observational emission features with synthetic EUV images generated from the model, and analyze their time evolutions. We also investigate the role of projection effects and line-of-sight integration in the measurements of plasma downflow speeds. Our analysis reveals that the Y-points do not necessarily coincide with the apparent cusp tip. Also, the apparent speeds of the supra-arcade downflows, as derived from tracks in the time--distance plots, underestimate the true Alfv\'{e}n speeds in the reconnection inflow region by at least a factor of 2 up to an order of magnitude. 

\end{abstract}




\section{Introduction}\label{sec:intro}

Solar flares are known as fast and explosive energy release phenomena, in which magnetic energy accumulated in nonpotential fields is suddenly released, causing immense plasma energization through heating, compression, and efficient acceleration of electrons and ions.
The standard flare model suggests that the explosive energy release is driven by magnetic reconnection occurring in the solar corona (see, e.g., \citealt{yamadaMagneticReconnection2010, shibataSolarFlaresMagnetohydrodynamic2011a,drakeMagneticReconnectionSolar2025} for review).

Observations of extended bright cusps in soft X-rays (SXRs), formed by a series of highly bent magnetic field lines in long-duration events (LDEs) well after the flare peak \citep{tsunetaObservationSolarFlare1992}, indicate that magnetic reconnection is not only occurring in a bursty fashion during the impulsive phase, but also can sustain the energy release in the gradual phase. 



Early works attempting to explain observed cusp shape and flare energy release \citep{tsunetaStructureDynamicsMagnetic1996} used the Petschek reconnection picture \citep{petschekMagneticFieldAnnihilation1964}. The observed visual tip of the cusp was interpreted to be associated with the primary reconnection X-point---a magnetic null point where magnetic connectivity breaks and field lines are reconnected to form closed loop configurations, allowing plasma inflow into the small-scale diffusion region and bidirectional high-speed outflows to either end of the latter. Attached to the diffusion region are slow-mode shocks that constrain the half-opening angle of the outflow jet to the dimensionless reconnection rate $\theta/2 \approx M \approx v_{in}/v_A$ (see \citealt{tsunetaStructureDynamicsMagnetic1996, longcopeEvidenceDownflowsNarrow2018a}). However, modern SXR and extreme ultraviolet (EUV) imaging shows that this region is full of intermittent plasma downflows and evolving ``fan'' emission in three dimensions, which is more consistent with an unsteady reconnection outflow region than a stable pair of Petschek slow shocks forming a neat wedge in the idealized 2D picture.

Observational evidence of the formation of large-scale current sheets during eruptive flares and coronal mass ejections  \citep[CMEs;][]{ciaravellaElementalAbundancesPostCoronal2002, linReviewCurrentSheets2015a} has revived the applications of theoretical models featuring a long and thin reconnection current sheet (RCS), based on Sweet--Parker \citep{sweetNeutralPointTheory1958a, parkerSweetsMechanismMerging1957} and Syrovatskii \citep{syrovatskiiFormationCurrentSheets1971} theories. In the idealized analytical picture, the current sheet bifurcates at both ends, where the current density is decaying \citep{Forbes1995, linEffectsReconnectionCoronal2000, bezrodnykhGeneralizedAnalyticalModels2011, forbesReconnectionPostimpulsivePhase2018}. The current sheet bifurcation sites, known as Y-points, are associated with local magnetic null points that mark the junctions between the RCS and the overlying and underlying closed magnetic field structures.

The quasi-laminar reconnection current sheet in the ideal model is unstable to various instabilities and becomes fragmented, potentially significantly altering the reconnection rate and magnetic geometry. 
Indeed, 2D numerical experiments that include resistivity and dissipation mechanisms have already shown that the current sheet is highly dynamic and structured \citep[e.g.,][]{shenNumericalExperimentsFine2011,shenDynamicalBehaviorReconnectiondriven2018,Mondal_2025}.
More recent 3D numerical experiments \citep{yeThreedimensionalSimulationThermodynamics2023, wangThreedimensionalTurbulentReconnection2023, ruanLorentzForceWork2024, renUnderstandingObservationalCharacteristics2025} demonstrate that the thickness of the current sheet may vary with height due to the expansion of reconnection outflows and associated turbulence. In addition, the presence of fine structures in the current sheet, such as magnetic islands and secondary reconnection sites, further complicates the topology, forming multiple microscopic X, O, and Y-points. In this scenario, the Y-points relevant to the macroscopic current sheet structure can not simply be regarded as the bifurcation point of the current sheet itself as in the idealized analytical 2D model (such as \citealt{linEffectsReconnectionCoronal2000}), but instead, need to be regarded 
as the locations with a general topological transition from quasi-parallel field geometry formed by highly bent field lines to closed magnetic loops in the ejecta or the flare arcade. 

Since the macroscopic Y-point is defined as a site of magnetic topology change, it implies that a definite determination of its position requires detailed knowledge of magnetic field geometry. However, the lack of vector coronal magnetic field measurements makes it challenging to make such a determination. 
One of the promising observational methods to determine the rapidly evolving magnetic field during flares is microwave imaging spectroscopy observations of gyrosynchrotron radiation from flare-accelerated energetic electrons \citep{garyMagnetographySolarFlaring2013}. Recently, using microwave data obtained by the Expanded Owens Valley Solar Array (\citealt{garyMicrowaveHardXRay2018}), a local minimum of the magnetic field strength is revealed in the above-the-loop-top cusp region \citep{chenMeasurementMagneticField2020}. This local minimum, referred to as the ``magnetic bottle,'' naturally arises near the bottom of the large-scale current sheet just above the closed flare arcade, and can be used to pinpoint the macroscopic Y-point  \citep{chenEnergeticElectronsAccelerated2024}. 
In the absence of direct magnetic field measurements, cusp-like features observed in EUV/SXR images---particularly in passbands sensitive to plasma temperatures of several million Kelvin and above---have frequently been used to infer the location of the macroscopic Y-point in eruptive flares \citep{suiEvidenceFormationLargeScale2003, liuReconnectingCurrentSheet2010, liuDynamicalProcessesVertical2013b, gouThermalPropertiesCurrent2024a}. This practice is motivated by the expectation that newly reconnected field lines retract and accumulate beneath the RCS, producing a hot, cusp-shaped morphology whose apex roughly traces the lower boundary of the reconnection outflow region. As a result, the apparent ``tip'' of the cusp is often taken as a proxy for the Y-point or the base of the reconnecting current sheet. However, this interpretation is inherently indirect and subject to potentially large uncertainties. EUV/SXR emission depends on plasma temperature, density, and line-of-sight integration, rather than magnetic topology alone, such that the visually identified cusp tip may reflect the region of strongest emission measure or heating rather than the true magnetic null. In addition, projection effects, foreground and background coronal emission, as well as inherent 3D structures, can all shift the apparent cusp apex relative to the actual reconnection geometry. Consequently, the location of the macroscopic Y-point inferred from EUV/SXR cusp morphology should be regarded as an approximate, emission-weighted indicator, rather than a precise tracer of the underlying magnetic geometry. 

\begin{figure}[ht!]
    \centering
    \includegraphics[width=0.95\columnwidth]{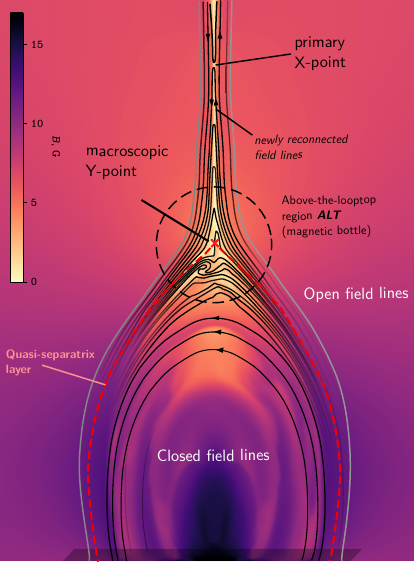} 
    \caption{Magnetic field geometry during the gradual phase of the solar flare, as derived from the numerical model \citep{shenOriginUnderdensePlasma2022}. Black lines indicate a streamline plot of the magnetic field for the X-Y slice at coordinate $z=0$. Background color map represents the magnitude of the magnetic field {at time step} $t = 3.0t_0$. The black dashed circle represents the above-the-loop-top region where a topological transition from quasi-parallel to closed geometry occurs.}
    \label{fig:ypt_cartoon}
\end{figure}

Another method to 
infer the positions of Y-points comes from the expected deceleration of plasma flows across the Y-point. 
Particularly, plasma downflows are expected to decelerate sharply near the lower Y-point, where they encounter the closed magnetic loops of the postflare arcade. Such a deceleration arises from the conversion of the downflows' kinetic energy into thermal and magnetic energy as the plasma is deflected and compressed by the strong magnetic and gas pressure gradients at the top of the flare arcade \citep{forbesNumericalExperimentRelevant1983, yokoyamaMagneticReconnectionCoupled1997, reevesCurrentSheetEnergetics2010}.
In observations, the speeds of the plasma downflows are usually measured using time--distance stack plots derived from time series EUV or SXR imaging observations. By placing a slit in the direction of the current sheet and the underlying flare arcade, one can visualize the plasma downflows as tracks with different slopes. A significant deceleration is usually found near the top of the flare arcade and has been interpreted as plasma flows passing the Y-point \citep{liuPLASMOIDEJECTIONSLOOP2013, yuMagneticReconnectionPostimpulsive2020, chenEnergeticElectronsAccelerated2024}.

The measurements of plasma downflow speeds 
not only serve as an important tool to constrain the location of the Y-points, but also may be used to quantify the energy release rate. Reconnection theories predict that the reconnection outflow speed is a substantial fraction of the Alfv\'en speed of the reconnecting magnetic field in the plasma inflow region $v_{\rm out}\approx v_{\rm A}^{\rm in}$. Using the observed downflow speed as a proxy for $v_A^{\rm in}$, and the observed speed of plasma flowing into the current sheet as the reconnection inflow speed $v_{\rm in}$, one may constrain the dimensionless reconnection rate as $M_A=v_{\rm in}/v_{\rm A}^{\rm in} $.

However, using this method, the reported reconnection rates in flares are often very large \citep{linDirectObservationsMagnetic2005, nagashimaStatisticalStudyReconnection2006, takasaoSimultaneousObservationReconnection2012, suImagingCoronalMagneticfield2013}, which sometimes approach or exceed the theoretical limit of $M_A^{\rm max}\approx 0.1$ \citep{YLiu2017,Yliu2022, cassakReview01Reconnection2017}.
One possibility that could account for this discrepancy is that some of the tracked plasma flows may not be the true reconnection outflows. It has been shown in observations that plasma flows above the flare arcade sometimes have a fast and slow population \citep{liuPLASMOIDEJECTIONSLOOP2013, yuMagneticReconnectionPostimpulsive2020}. The former, which has speeds of $\sim$200--1000~km~s$^{-1}$, is usually located at high altitudes, and may be more representative of the true reconnection outflows. The latter, which is considerably slower with speeds of $\lesssim$100~km~s$^{-1}$, may represent the already slowed-down plasma flows or contracting loops below the Y-point. These plasma downflows, sometimes referred to as supra-arcade downflows (SADs) or supra-arcade downflowing loops \citep{McKenzie1999,savageRECONNECTIONOUTFLOWSCURRENT2010}, have been argued to be the signatures of the wakes of the contracting flux tubes \citep{savageREINTERPRETATIONSUPRAARCADEDOWNFLOWS2012} or complex plasma flows developed in the turbulent ``interface region'' where fast reconnection outflows impinge upon the underlying flare arcade \citep{shenOriginUnderdensePlasma2022}. In either case, a subset of the observed plasma downflows may be substantially slower than the true reconnection outflows. Consequently, when using their speeds as the proxy for $v_{\rm out}$, the inferred reconnection rates may be a factor of a few to nearly an order of magnitude greater than the true values.

Another caveat is that the speeds of the reconnection outflows $v_{\rm out}$ may also be nonuniform and can be substantially smaller than the Alfv\'{e}n speed in the inflowing region $v_{\rm A}^{\rm in}$ itself. Numerical simulations and \textit{in situ} measurements in space plasmas indicate that even within the exhaust region, the bulk outflow speed typically reaches only a fraction of the upstream Alfvén speed, around 0.3--0.8$v_{\rm A}^{\rm in}$, rather than the ideal Alfvénic value expected from simplified reconnection models \citep[e.g.,][]{Paschmann1986,Linton2006,Haggerty2018}. This reduction has been attributed to finite guide fields, pressure anisotropy, and energy partition into thermal heating and instabilities within the exhaust. Moreover, once reconnection outflows exit the diffusion region (extremely small in the solar flare context), additional deceleration can occur due to interaction with the ambient plasma. It has been suggested that an aerodynamic drag force may be responsible for slowing down the reconnection outflows \citep{savageQUANTITATIVEEXAMINATIONLARGE2011}, which was subsequently modeled by \citet{Unverferth2021} to compare with observations of the 2017 September 10 flare \citep{longcopeEvidenceDownflowsNarrow2018a,yuMagneticReconnectionPostimpulsive2020}. Also, numerical experiments have shown that structures in the reconnection outflow region, such as macroscopic plasmoids, can also be substantially slower than the surrounding outflows by as much as 80\% \citep{shenNumericalExperimentsFine2011}. As a result, speeds inferred from remotely observed plasma motions, even if they represent true flow dynamics in the RCS, may substantially underestimate the intrinsic Alfv\'{e}n speed in the reconnection inflow region, which, in turn, leads to an overestimate of the reconnection rate.

Finally, it could be the case that the indirect velocity measurement inferred from time--distance diagrams derived from the EUV/SXR time series images may not truthfully represent the highest achievable speeds of the reconnection outflows. In fact, due to the projection and integration of all the complex dynamic and fine structures along the line of sight, the derived speeds may be skewed to lower values by the denser and relatively slower-moving structures in the current sheet. 

Taking advantage of a recently developed 3D numerical model of flare reconnection by \citet{shenOriginUnderdensePlasma2022}, 
our study explores the role of line-of-sight integration effects in (a) localizing the macroscopic reconnection Y-point by using the observed cusp structure and (b) interpreting the observed speeds of the plasma downflows using the time--distance plot method.
We will use the 3D model to produce temporally resolved synthetic EUV and SXR images, and compare the results to observations of the prominent candle-flame-shaped solar flare of 2012 July 19, with a focus placed on the cusp tip location and plasma downflow speeds. Section~\ref{sec:model} briefly introduces the 3D magnetohydrodynamics (MHD) model and an example flare event. In Section~\ref{sec:morphology}, we discuss the identification of the macroscopic Y-point in the numerical model and compare the morphology of the cusp-shaped structure in the synthetic and observed EUV/SXR images. In Section~\ref{sec:evolution}, we investigate the apparent speeds of the plasma downflows 
acquired from the synthetic time--distance maps and find that they strongly underestimate the 
actual speeds of the reconnection outflows. We then summarize and discuss the implications in Section~\ref{sec:conclusion}.




\begin{figure*}[ht!]
    \centering
    \includegraphics[width=0.95\textwidth]{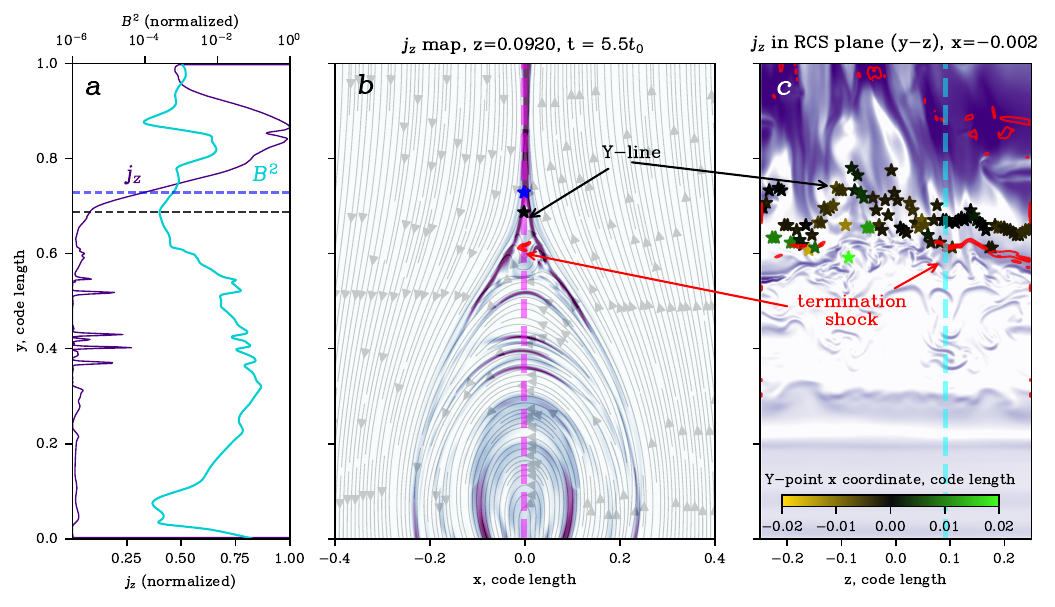}
    \caption{RCS geometry and magnetic field configuration at a selected time, $t = 5.5t_0$, during the MHD model evolution.
    Central panel (b) shows the two-dimensional distribution of 
    current density $j_z$ calculated for the single $x$--$y$ slice of the simulation dataset. Overlaid in light gray are the streamlines of the magnetic field derived in the cutting plane. {Blue and black markers show} the bifurcation of the current sheet and the local magnetic field minimum closest to the bifurcation site.
    {The pink dashed vertical line in panel (a) marks the position of the RCS and defines the $x$-coordinate of the $y$--$z$ plane, where the cut across the RCS was taken, shown in panel (c). The cyan dashed vertical line in panel (c) shows the $z$-coordinate where the $x$--$y$ slice shown in panel (a) was taken.}
    The red patch shows the contours of the highest velocity convergence $\max( - \nabla\cdot \mathbf{v})$ that is expected to trace the position of the termination shock {in the $y$--$z$ plane}. {Panel (a)} shows the $j_z$ distribution along the $z$-axis extracted from the vertical slice along the RCS (shown in the central panel).
    {Panel (a)} shows local current density $j_z$ (purple) and magnetic field $B^2$ profiles (cyan) extracted along the RCS. A dashed blue line corresponds to the height of the bifurcation point, while the black line highlights the local magnetic field minimum closest to the RCS bifurcation site.
    {Panel (c) shows the associated current density distribution $j_z$ computed along the $y$--$z$ slice chosen along the vertical dashed line shown on the central panel. Stars trace the position of the Y-points in the $z$-direction identified from the RCS bifurcation and local magnetic minimum. The color indicates the displacements of Y-points along the $x$ direction.}
    }
    \label{fig:ypts}
\end{figure*}


\section{Overview of the 3D MHD Model and Flare Event}\label{sec:model}
To gain insights into the three-dimensional dynamics of reconnecting magnetic fields in the flare current sheet and the above-the-loop region, we analyze the results of a resistive 3D MHD model developed by \citealt{shenOriginUnderdensePlasma2022}. More detailed descriptions of the model can be found in \citet{shenOriginUnderdensePlasma2022}. Here we only provide a brief overview. The 3D MHD model is initialized from a 2.5D model of reconnection in the solar corona based on a standard Harris current sheet with the lower boundary of the simulation domain characterized by photospheric density ($n_e=10^{12}$~cm$^{-3}$) and line-tied magnetic field at the bottom, which further leads to the formation of closed post-flare loops or Kopp--Pneumann configuration once the magnetic reconnection occurs within this vertical current sheet following an initial perturbation. After the closed postflare loops are formed in a 2.5D setup, physical parameters are extended uniformly along the third dimension, and the simulation is run self-consistently in 3D, which leads to the development of new reconnection sites within the current sheet and formation of plasmoids, as well as generating turbulence and plasma instabilities. {Since the classic flare geometry is well established in the 2.5D model, the approach extending from 2.5D to 3D allows us to explore the flare cusp region in the classic flare framework. However, it is noted that this method does not include an eruption nor a full 3D representation of plasma evolution along the guide-field direction ($z$-axis). Future work will require full 3D simulations to address these system-scale dynamics.}
The model includes treatments of anisotropic thermal conduction, approximation of coronal heating function, and radiative cooling. Such treatments are crucial for a proper description of thermodynamical processes that involve intense plasma heating and cooling in the above-the-loop-top region \citep{yokoyamaMagneticReconnectionCoupled1997}, which are necessary for producing realistic EUV and SXR time series images and comparing them to the observations.

The plasma conditions in the initial run of the MHD simulation are defined with the Lundquist number $S=5\times 10^4$ and plasma $\beta=0.1$ in the ambient coronal plasma. The characteristic physical parameters that were used to normalize MHD equations to the nondimensional form correspond to the typical values in the solar corona: length $L_{\text{char}}=150$~Mm, magnetic field $B_{\text{char}}=10$~G, density $n_0=2.5\times 10^8$~cm$^{-3}$, temperature $T_0=1.13 \times 10^8$~K, Alfv\'en speed $V_{\rm A}=1366$~km/s and time scale $t_0=109.8$~s. Simulation was performed using a 3D uniform Cartesian grid with $[579, 576, 288]$ cells along $x, y,$ and $z$. The simulation domain is defined as $-0.5 \leqslant x \leqslant 0.5$, $0 \leqslant y \leqslant 1$, and $-0.25 \leqslant z \leqslant 0.25$ in code units.

Figure~\ref{fig:ypt_cartoon} shows the magnetic field configuration after the reconnection in the current sheet has been established, which corresponds to the gradual phase of an LDE that hosts a bright candle-flame-shaped feature. The time step of interest is selected in a way that the magnetic field in the above-the-loop-top region would undergo a steady three-dimensional evolution by developing plasma instabilities and forming a wavy, irregular reconnection X-line, thus resulting in a configuration substantially distinct from the initially symmetric extension from the two-dimensional setup.

To perform a model-to-data comparison, we use observations of a well-studied long-duration eruptive solar flare that occurred on 2012 July 19. This flare occurred on the southwestern region of the solar limb and provides a favorable viewing geometry with the current sheet viewed nearly edge-on. It allows us to study the candle-flame-shaped morphology and plasma downflows with minimal impediment caused by projection effects due to event orientation. 
Bright SXR and EUV emission from the cusp feature above the postflare arcade is observed over the course of almost 12 hr during the gradual phase (from 06:00 UT until about 18:00 UT), which featured enduring high-altitude plasma downflows. For a detailed summary of EUV and X-ray observations of the impulsive and gradual phase of the event, we refer interested readers to other papers \citep{kruckerPARTICLEDENSITIESACCELERATION2013b, liuDynamicalProcessesVertical2013b, liuPLASMOIDEJECTIONSLOOP2013}. In this work, we select several 20~minute time intervals during the gradual phase, considerably long after the flare impulsive phase, when the associated CME (also mentioned as the erupted flux rope) has propagated to a large distance. The time selection is to ensure that the geometry of the flare aligns with our model assumptions, which feature a long current sheet with an underlying reconnected flare arcade system, but without a flux rope. 

We use observations from the Atmospheric Imaging Assembly (AIA) onboard the Solar Dynamics Observatory (SDO; \citealt{lemenAtmosphericImagingAssembly2012}) in the 94~and 131~\AA\,channels, which contain the main contribution from the Fe~$XVIII$ and Fe~$XXI$ emission lines with temperature response peaking at 6.3~MK and 10~MK, respectively \citep{odwyerSDOAIAResponse2010, zannaMultithermalEmissionSolar2013}. SDO/AIA data are available over the entire duration of the event at a time cadence of 12 s.
Full-disk images are calibrated and processed using the standard \texttt{aia\_prep} routine (updated pointing and registration) included in the \texttt{SunPy} analysis software \citep{sunpycommunitySunPyPythonSolar2015, barnesAiapyPythonPackage2020}. 

Imaging observations by the X-ray Telescope (XRT) on board the Hinode spacecraft (\citealt{golubXRayTelescopeXRT2008}) instrument with the Open-Al~thick filter pair are the closest in resembling the candle-flame brightness distribution observed in the AIA 131~\AA\,channel. The Al~thick filter has a peak of its temperature response function in a similar temperature range as AIA 131~\AA, at $T\sim10$~MK. XRT observations are calibrated to Level 1 using the \texttt{xrt\_prep} procedure available in the {\texttt{SolarSoft} IDL package (\texttt{sswIDL})} \citep{freelandDataAnalysisSolarSoft1998}. The XRT images are normalized to display output in units of DN per second. To perform joint analysis with AIA observations, the coalignment routine \texttt{xrt\_read\_coaldb} is used to update instrument coalignment using a reference AIA 335~\AA\,image and XRT coalignment database\footnote{\url{http://ylstone.physics.montana.edu/yosimura/hinode/coalignment/}} to apply rotation corrections to the XRT image.

\begin{figure*}[!p]
\centering
\includegraphics[width=1.0\textwidth]{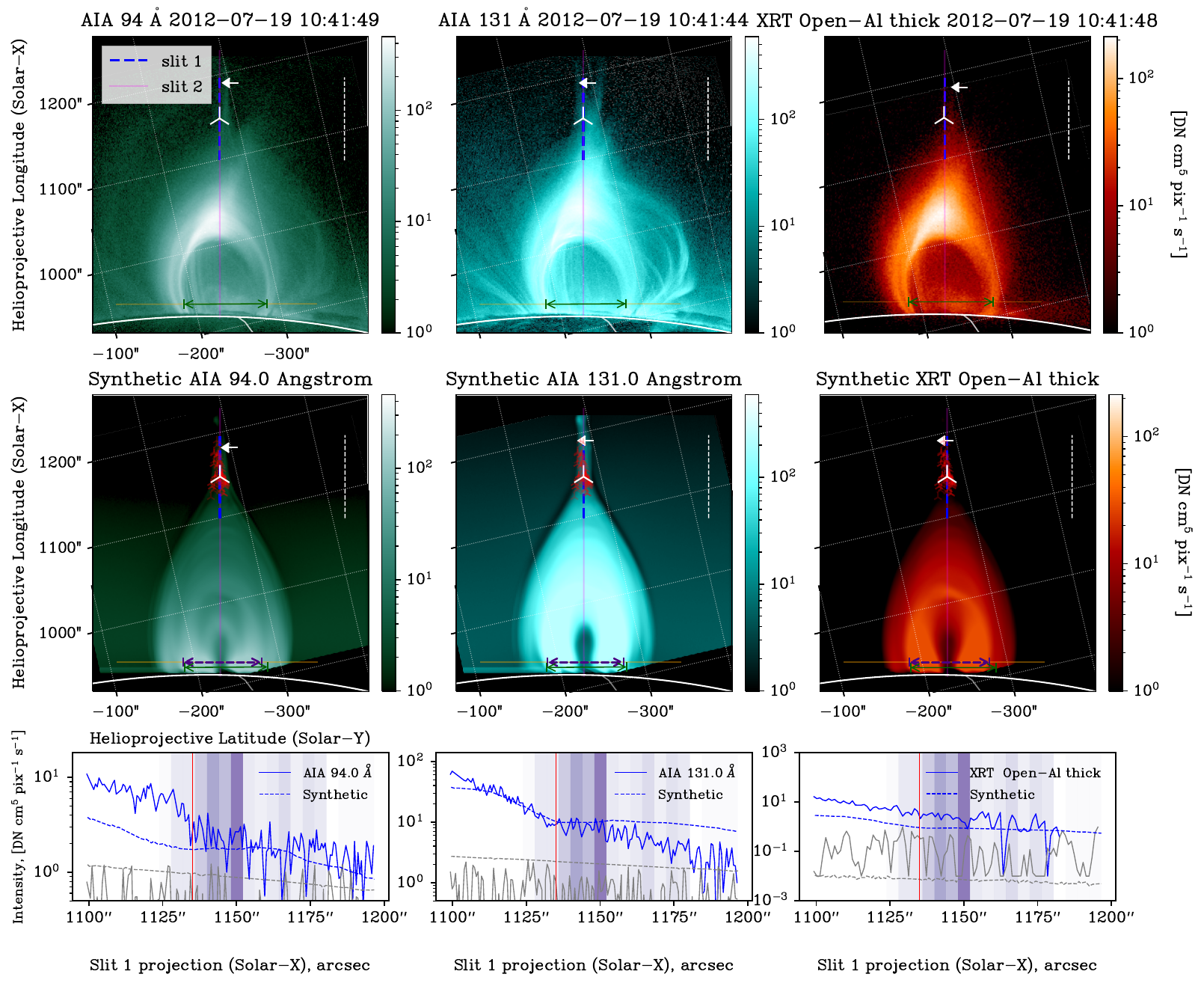} 

\caption{Observed and modeled EUV images of the candle-flame-shaped flare that occurred on 2012 July 19. The top three panels, from left to right, show images in the high-temperature EUV (94 \AA, 131 \AA) and SXR (Al thick) channels obtained by SDO/AIA and Hinode/XRT. Observation time was chosen in the gradual phase of the flare at 10:41 when data from XRT is available. The middle panel row presents synthetic images computed from the 3D MHD simulation dataset by \cite{shenOriginUnderdensePlasma2022} using temperature response functions for corresponding channels in AIA and XRT detectors. Red markers indicate projected positions of Y-points as derived from the numerical model. An upside-down white ``Y'' marker indicates the mean position of the model-derived Y-points projected on the vertical slit \#1. White arrows indicate the location of the ``visual'' tip of the cusp based on a 5$\sigma$ level relative to the coronal background. {Double-headed arrows mark the approximate widths of postflare loops for observed (green solid) and synthetic images (purple dashed) as derived from brightness profiles extracted along orange horizontal slits.}
Bottom panel features synthetic (dashed) and observational (solid) brightness profiles extracted along the vertical slit \#1 aligned with the position of the RCS in the numerical model (blue dashed lines). Slit \#1 is centered at the position of the macroscopic Y-point. {The white dashed line placed close to the right edge of the image indicates the slit used to extract the brightness profile in the background coronal region.} Coronal background brightness profiles (gray) {shown on the bottom panels} are extracted along the white dashed slit. Plotted in purple is the histogram showing the distribution of projected Y-points on the slit plane.  {Vertical red line in each of the bottom panels indicates where the transition from exponential decay to plateau occurs.}}
\label{fig:obs_slits}
\end{figure*}

\section{The Reconnection Y-Points}\label{sec:morphology}




In this section, our goal is to examine how we can use the observed cusp geometry in EUV and SXR imaging to constrain the site where the transition between the 
RCS and closed magnetic field line configurations occurs, namely, the lower reconnection Y-point. This is first done by identifying the location of the Y-points based on the known magnetic field configuration in the simulated data. Then, by synthesizing EUV and SXR emission from the model, we compare the morphology of the apparent cusp feature to the identified locations of Y-points. Such a comparison, in turn, allows us to inform the interpretation based on the observations.

To generalize the 2D concept of a Y-point in the ideal Sweet--Parker reconnection model to more realistic 3D resistive numerical simulations, additional criteria to locate the macroscopic Y-points in the 3D MHD model are required. 
In our study, we reconstruct a three-dimensional Y-line in the simulation output by locating a Y-point inside each 2D slice extracted at different $z$ coordinates along the RCS.
For each slice, the identification of the lower macroscopic Y-point at the bottom of the current sheet is done by using two complementary methods: (1) finding where the current sheet shows a bifurcation; (2) identifying magnetic null points, which exhibit as the local minima of the magnetic field strength~$|B|$. The current sheet bifurcation is due to the transition in topology from nearly anti-parallel field lines in the current sheet to cusp-shaped, closed field lines above the flare arcade. The local minimum of $|B|$, or the magnetic bottle, is due to the field line divergence at this location (see Fig. \ref{fig:ypts}(a), (b)), which naturally arises due to the presence of a large-scale current sheet above the flare loop top (see \citealt{chenEnergeticElectronsAccelerated2024} for a more detailed discussion). To find the bifurcation site, we identify a branch point of the skeletonized structure of the current density distribution in each 2D slice (in the $x\text{-}y$ plane). We further mark the local magnetic minimum, closest to the bifurcation site, as the Y-point. Details about the Y-point recognition algorithm are given in Appendix \ref{app:yptsrec}.

{Figure~\ref{fig:ypts}(c) shows the distribution of identified Y-points from 90 $x\text{-}y$ slices extracted at different $z$ coordinates. We show positions of identified Y-points in the $y-z$ plane with color-coding indicating displacement along the $x$-axis. The identified Y-points are localized below the bottom of the current sheet, seen as a sharp transition in $j_z$, and above the termination shock. Sometimes, fine structures such as plasmoids (or magnetic flux tubes) could spontaneously develop within the current sheet and modulate the fine structures near the Y-point.  
The evolution of the Y-points at all simulation time steps is shown in Fig.~\ref{fig:ypt_animation} (see Appendix~\ref{app:ypt_rcs_scan}). It demonstrates that the perturbation of the current sheet structures could result in a complex spatiotemporal distribution of the Y-points.}

To carry out model observation comparisons, we first determine the viewing geometry by matching the orientation of the observed postflare arcade.  This is done by combining multiperspective observations by SDO/AIA and the EUVI instrument on board the STEREO observatory \citep{wuelserEUVISTEREOSECCHIExtreme2004} in the late gradual phase using lower temperature ($T\sim 1.5$~MK) 193~\AA~and 195~\AA~channels, respectively. {We infer the angle $\theta$} between the AIA line of sight and the plane of the current sheet to be 5$^{\circ}$--10$^{\circ}$ {using the observed orientations of several postflare EUV loops shown in Figure~\ref{fig:alignment}}. This angle may contribute, at least in part, to the apparent thickness of the hot plasma spikes observed by AIA and XRT at the tip of the cusp, {and the width of the postflare arcade in the synthetic images}.
The details of the alignment procedure and coordinate transformations are given in the Appendix \ref{app:align}.

\begin{figure*}[ht!]
\centering
\includegraphics[width=1.0\textwidth]{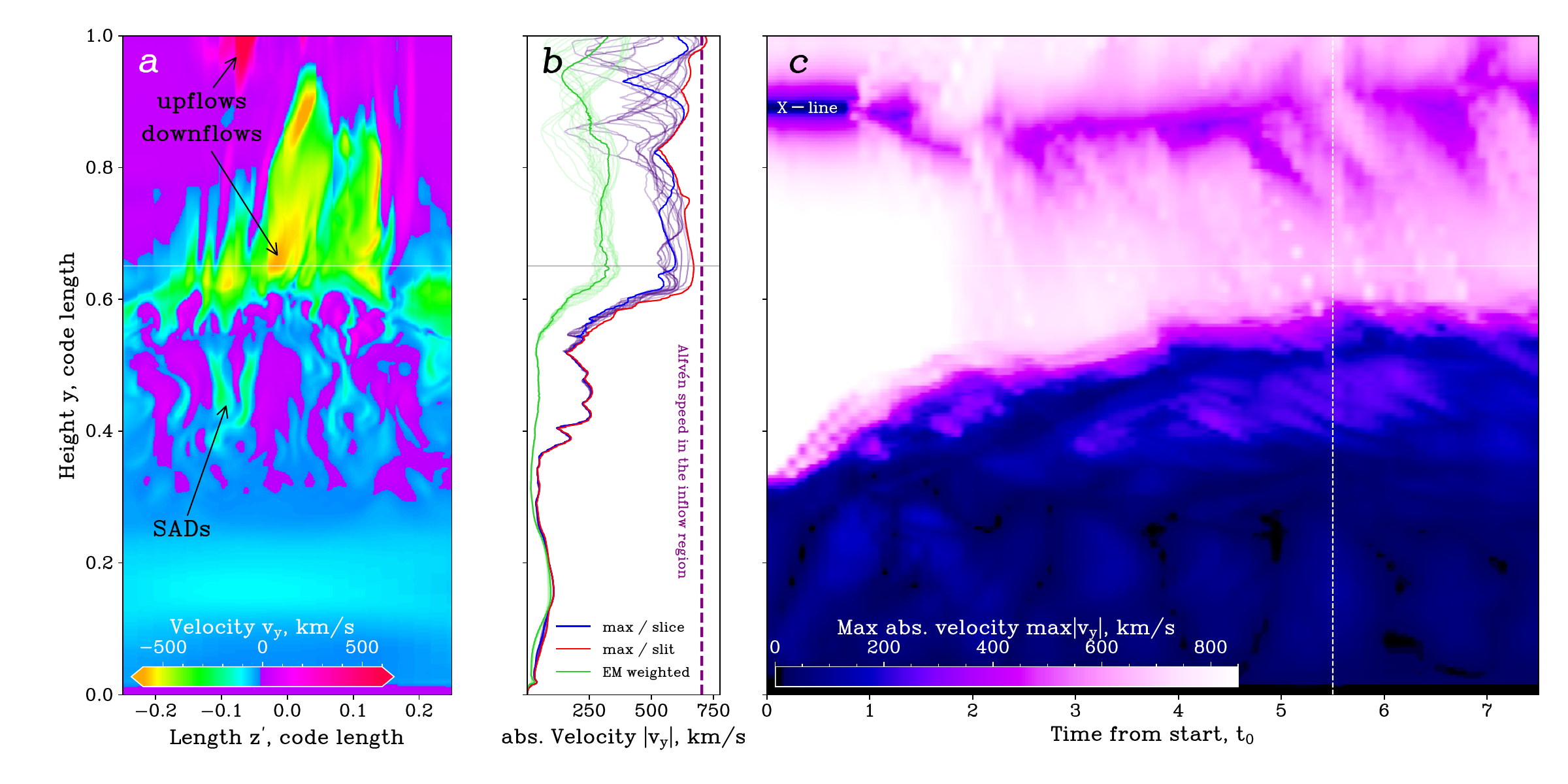} 
\caption{Distribution of plasma flow velocities derived for an oblique slice through the plane of the RCS.
Panel (a) shows an oblique slice through the 3D MHD dataset (along $z'$) that was chosen to be aligned with the slit \#2 defined in Fig. \ref{fig:obs_slits}. This slice shows the distribution of the vertical component of plasma velocity $v_y$ and illustrates the geometry of plasma flows inside the RCS.  
The central panel (b) shows EM-weighted $\langle v\rangle_{EM}$ (light green) and maximum velocity profiles (purple) with height for each one of 20 2D slices parallel to the slice \#10 (shown on panel (a)).
Blue profile shows the maximum velocity at each height calculated for slice \#10. The red line is an envelope curve calculated from all maximum velocity profiles extracted from each slice.
Maximum $v_y(y)$ profiles of plasma flows in the slit plane computed for every simulation timeframe were stacked into a single time--distance diagram shown on the right panel (c). The color map represents the maximum instantaneous plasma velocity of the reconnection downflow $v_y(y, t)$ picked along the line of sight to identify individual downflows.
}
\label{fig:velmap}
\end{figure*}

From this viewing perspective, we further compute synthetic EUV and SXR emissions from the plasma parameters available in the 3D MHD model and create synthetic emission maps using the \texttt{rushlight} package\footnote{\url{https://rushlight.readthedocs.io/}}. To derive synthetic EUV/SXR brightness distributions, for each location in the plane of the sky ($x',y'$), we first compute the EUV and SXR emissivity in each simulation cell $dI_i=n_i^2R_m(T_i)dz'_i$, where $n_i$ and $T_i$ are the plasma density and temperature of each cell $i$ located along the line of sight (denoted as the $z'$ axis after rotating the 3D simulation cube), $dz'_i$ is the projected grid size in the line-of-sight direction, and $R_m(T_i)$ is the instrumental response function of the selected filter $m$ for a given instrument (SDO/AIA or Hinode/XRT). 
For optically thin EUV and SXR emission, the final intensity at the ($x',y'$) location is the direct summation of all contributions from these cells along the line of sight
\begin{equation}
    I_{m}(x',y') = \sum_i dI_i(x',y') = \sum_i n_i^2 R_m(T_i) dz'_i.
\end{equation}
When deriving temperature response functions using \texttt{aia\_get\_response.pro} and \texttt{make\_xrt\_temp\_resp.pro} routines in \texttt{sswIDL}, we assume ionization equilibrium and the standard coronal abundance model from the \texttt{CHIANTI} database \citep{dereCHIANTIAtomicDatabase1997, delzannaCHIANTIAtomicDatabase2021}.

Synthetic images are projected into the observer's image plane and aligned using the derived position of a postflare loop midpoint location (from inter-footpoint distance) and the current observer's coordinate derived from the SDO/AIA FITS file header. The image in each synthetic SDO/AIA and Hinode/XRT filter is then resampled into the corresponding pixel grid of the reference image to the corresponding pixel scale for AIA and XRT (0.6~\arcsec pix$^{-1}$ for AIA and 1.02~\arcsec pix$^{-1}$ for XRT).

Figure~\ref{fig:obs_slits} shows the results of our model-to-data comparisons for AIA 94~\AA\ (left column), AIA~131~\AA\ (middle column), and XRT~Al~thick (right column). The synthetic images successfully reproduce most main features of the observations, including the correct intensity level and the cusp-shaped flare arcade in all three EUV/SXR filter bands.  
The best agreement is observed for AIA/131\AA~channel, since the original version of the model \citep{shenOriginUnderdensePlasma2022} was designed to reveal the dynamical structures in high-temperature current sheets and flare loop-top regions ($\sim 10^7$~K) associated with fast magnetic reconnection. 
We note that, however, the postflare arcade in the synthetic images is notably different from the observations. Its bottom portion appears brighter and wider, while the loop top lacks a prominent enhancement, as in the observations. Reproducing the realistic brightness of the closed loops at lower coronal heights would require more rigorous treatment of radiative cooling and realistic chromospheric evaporation driven by accelerated particles, which are not included in this model.  
Another morphological difference lies in the slight asymmetry seen in the observed cusp feature, which is not reproduced in the synthetic images because our model is set to start from a symmetric reconnection geometry. 

After calculating the alignment and scaling of the MHD model for the 2012 July 19 event, we re-project the 3D coordinates of the identified Y-points positions in the model on the synthetic images. 
The distribution of the derived Y-points is shown in Figure~\ref{fig:obs_slits}.
Most of the projected Y-points correspond to those along a continuous Y-line; therefore, 
we estimate the mean position of the projected cluster of Y-points in the plane of sky ($x', y'$), which would represent a ``macroscopic'' Y-point (shown as a white upside-down ``Y'' on Fig. ~\ref{fig:obs_slits}).

 

Next, we examine how these Y-point locations compare with the apparent ``tip'' of the cusp feature. We note that the visual estimate of the tip location is inherently subjective, which is very sensitive to the color map/color scale selection and, additionally, affected by the dynamic range and background noise level of the EUV/SXR instruments. To make such a determination more objective, we determine the location of the visual ``tip'' based on the coronal background level extracted outside of the cusp structure along the slits shown as white dashed lines on Fig. \ref{fig:obs_slits}. We set the lower limit for the brightness of the ``tip'' to be at 5 times the rms of the background corona $\sigma$, {to confidently separate the brightness levels within the current sheet region from the ambient coronal levels and noise inherent to AIA and XRT measurements in the background regions} (as shown by white arrows on Fig. \ref{fig:obs_slits}). 
It can be seen that for all synthetic images, the distributions of the projected Y-points are clustering well below the visual tip. In fact, the extended spike features in the cusp region, or extensions of cusp peaks to higher altitudes (as can be seen in the observed AIA 131\,\AA~image) may represent the lower portion of the RCS, but not necessarily the newly reconnected highly bent loops below the macroscopic Y-point.

To further elucidate the relative connection between the Y-point locations and the observed EUV/SXR intensity distribution, we analyze brightness profiles extracted along a vertical slit and overplot the histogram of the projected Y-point coordinates on the slit. While the Y-points are distributed in a variety of heights, the majority of them are located in the transition region from a steep exponential drop of the intensity profile toward a ``plateau'' that eventually merges into the coronal background at larger heights. 

From the MHD model, we see that the sections of the brightness profile featuring the steep exponential drop and the shallow plateau correspond to the closed flare arcade and the RCS regions, respectively. The steep exponential profile in the closed flare arcade is due to the strong tension force exerted on the flare arcade plasma by the highly bent, closed loops below the Y-point. This extra downward force strongly compresses the plasma, resulting in a steep exponential density profile with a small density scale height. In contrast, in the RCS region above the Y-point, the plasma $\beta$ is high. As such, the magnetic tension force there does not play a key role, and the density gradient is mainly governed by the hot, $\sim$10~MK plasma in a quasi-equilibrium state, which sets a much greater density scale height and hence a less steep density profile.

Thus, compared to using the visual tip of the cusp, the transition region between these two brightness profile regimes may be a better indication of the location of the macroscopic Y-point based on the EUV/SXR images.

\section{Plasma Downflow Speeds}\label{sec:evolution}
\subsection{Plasma downflows inside the RCS}\label{subsec:dwflid}

As introduced earlier, measurements of plasma downflow velocities from the apparent motions of 
brightness features in the EUV and SXR time series images using the time-distance method could be confounded by projection effects and the complicated interplay of multiple overlapping plasma flows along the line of sight. In this section, we use the 3D MHD model to investigate these effects.

\begin{figure*}[ht!]
    \centering
    \includegraphics[width=0.8\textwidth]{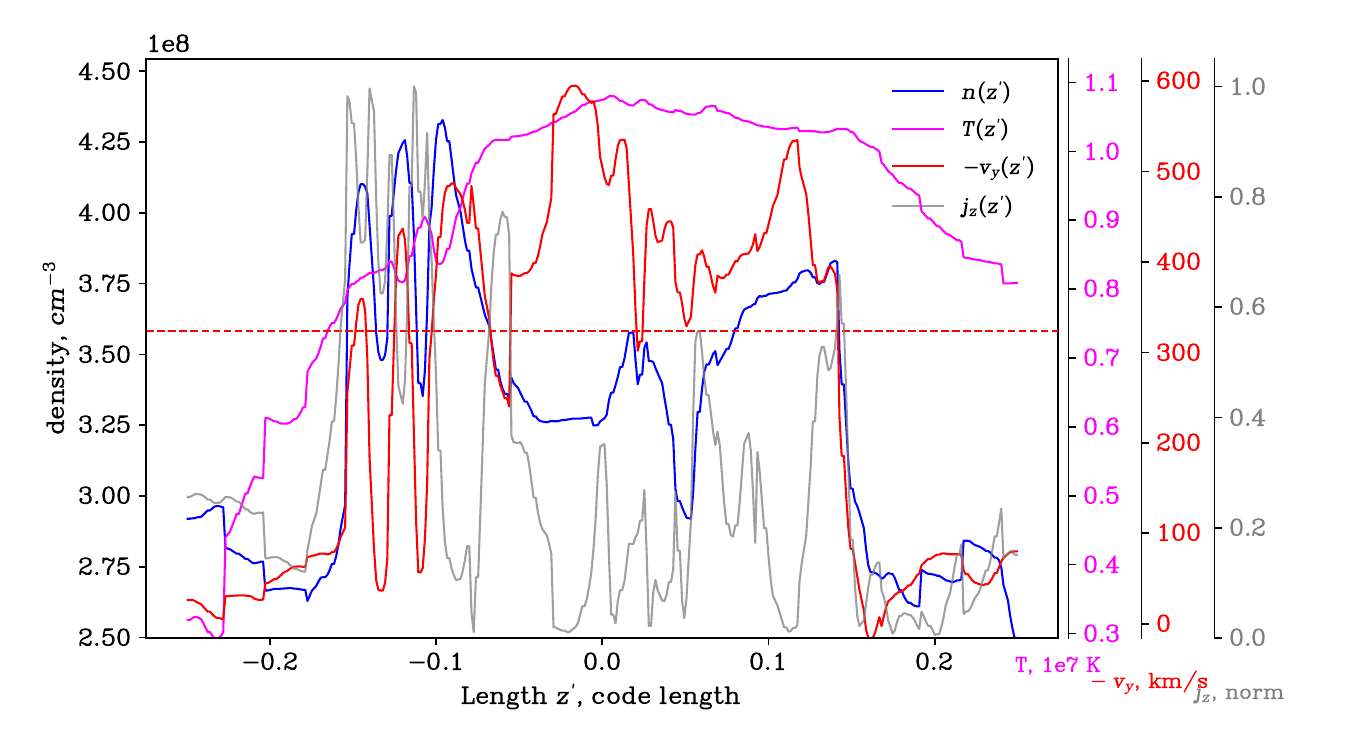} 
    \caption{Plasma parameters distribution along the line of sight determined for the horizontal slice along $z'$ shown in Fig.~\ref{fig:velmap}. The red line shows the vertical velocity $v_y$ inside the slit plane. Blue and magenta lines correspond to the density and temperature within the current sheet, respectively. Gray profile represents the normalized current density $j_z$ profile and determines the extent of the RCS intersection with the slit plane. The red dashed line indicates EM-weighted velocity $\langle v\rangle_{\rm EM}$.}
    \label{fig:param_dist_los}
\end{figure*}


\begin{figure*}[ht!]
\centering
\includegraphics[width=1.0\textwidth]{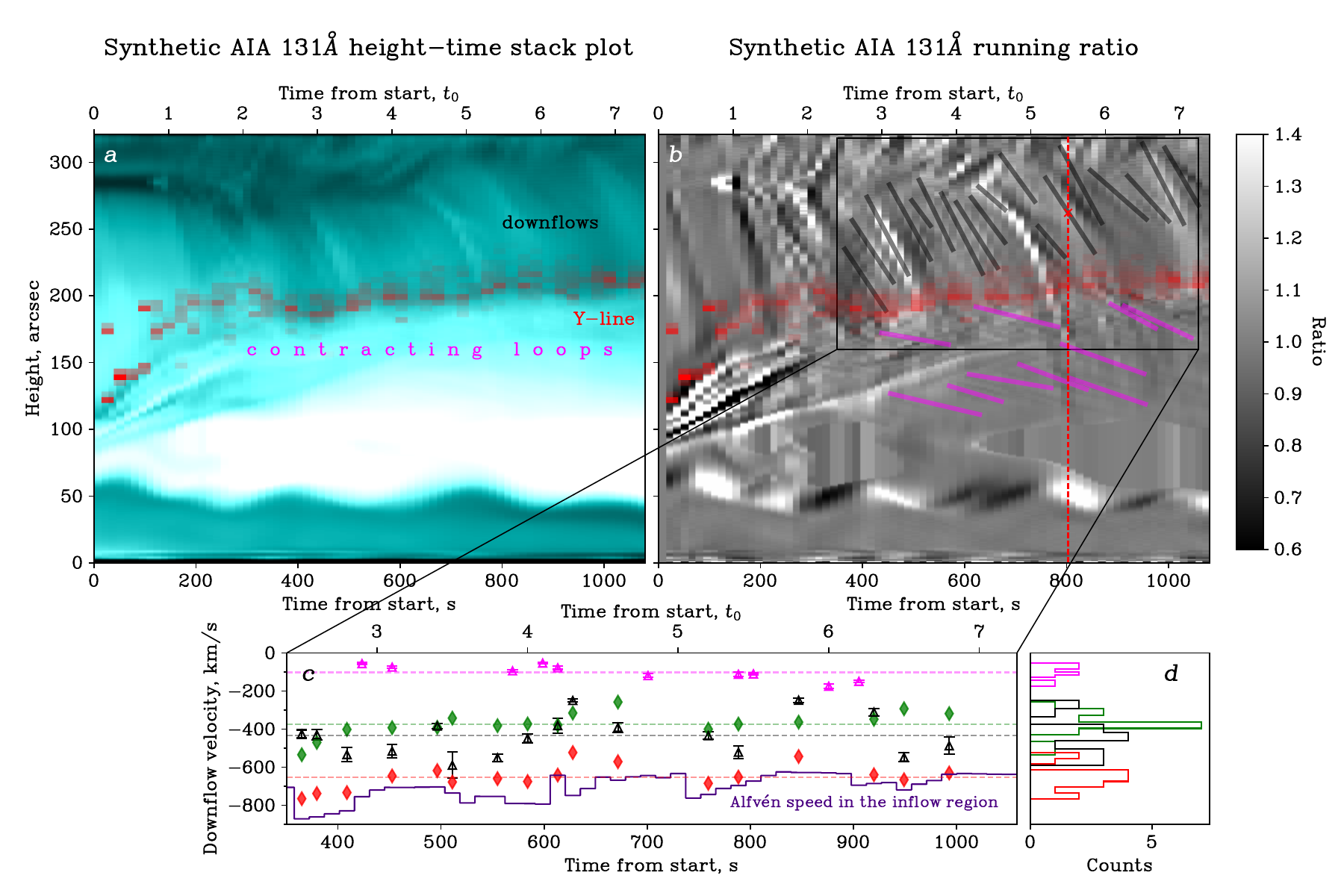}
\caption{Comparison between plasma flow speeds estimated from height--time stack plots and the model-derived maximum and EM-weighted speeds.
Panel (a) features a height--time stack plot of synthetic AIA 131~\AA\,brightness extracted along the slit~\#2 defined in Fig. \ref{fig:obs_slits}. The red shading illustrates the histogram of Y-point positions as projected on the slit for each timestep of the simulation.
(b) Time--distance diagram computed from the running ratio of synthetic AIA 131~\AA~images featuring fast plasma downflows in the upper part of the plot (dark gray lines), and slow plasma downflows (magenta lines) corresponding to contracting loops.  The red vertical line shows the time step analyzed previously in Figures \ref{fig:ypts}, \ref{fig:obs_slits}, \ref{fig:velmap}. The lower panel (c) shows inferred velocities of the fast and slow plasma downflows from the linear fits (black and magenta triangles). Green diamonds indicate the EM-weighted speed of all fast downflows intersecting the selected line of sight at a given height. Red diamonds indicate the maximum speed of the downflow that intersects the line of sight. 
The purple line indicates the value of {local Alfv\'en speed calculated in the vicinity of the reconnection X-point, seen as the origin of bidirectional outflows at a height of about 300~\arcsec in panels (a) and (b)}.
Histogram on panel (d) shows maximum (red), EM-weighted (green), and time--distance--derived (black) speed distributions of all analyzed downflows.
}
\label{fig:dwfls}
\end{figure*}


First, to characterize the velocity distribution of the plasma downflows within the RCS, we extract the instantaneous vertical velocity distributions ($v_y$) in the oblique slit plane that crosses the RCS in the MHD model under an angle of $\theta\sim~5^{\circ}$ (as shown in Fig. \ref{fig:alignment}). 
Figure~\ref{fig:velmap} shows the velocity distributions in the slit plane, featuring reconnection downflows originating at the $X$-line in the upper part of the simulation domain at $y \sim 0.9$, and impinging on the postflare arcade at $y\sim 0.6$. From the velocity distribution in the $y\text{-}z'$ plane (Figure~\ref{fig:velmap}(a)), we can tell that the fastest downflows with typical $v_y\sim~-700$~km/s appear in the center of the oblique slice that intersects with the center of RCS at $z'\sim0$ (or $x=0$).
These fastest plasma downflows are enclosed by the background coronal plasma that has a dramatically lower $v_y$ component (see Fig. \ref{fig:param_dist_los}).   




\begin{figure*}[ht!]
\centering
\includegraphics[width=1.0\textwidth]{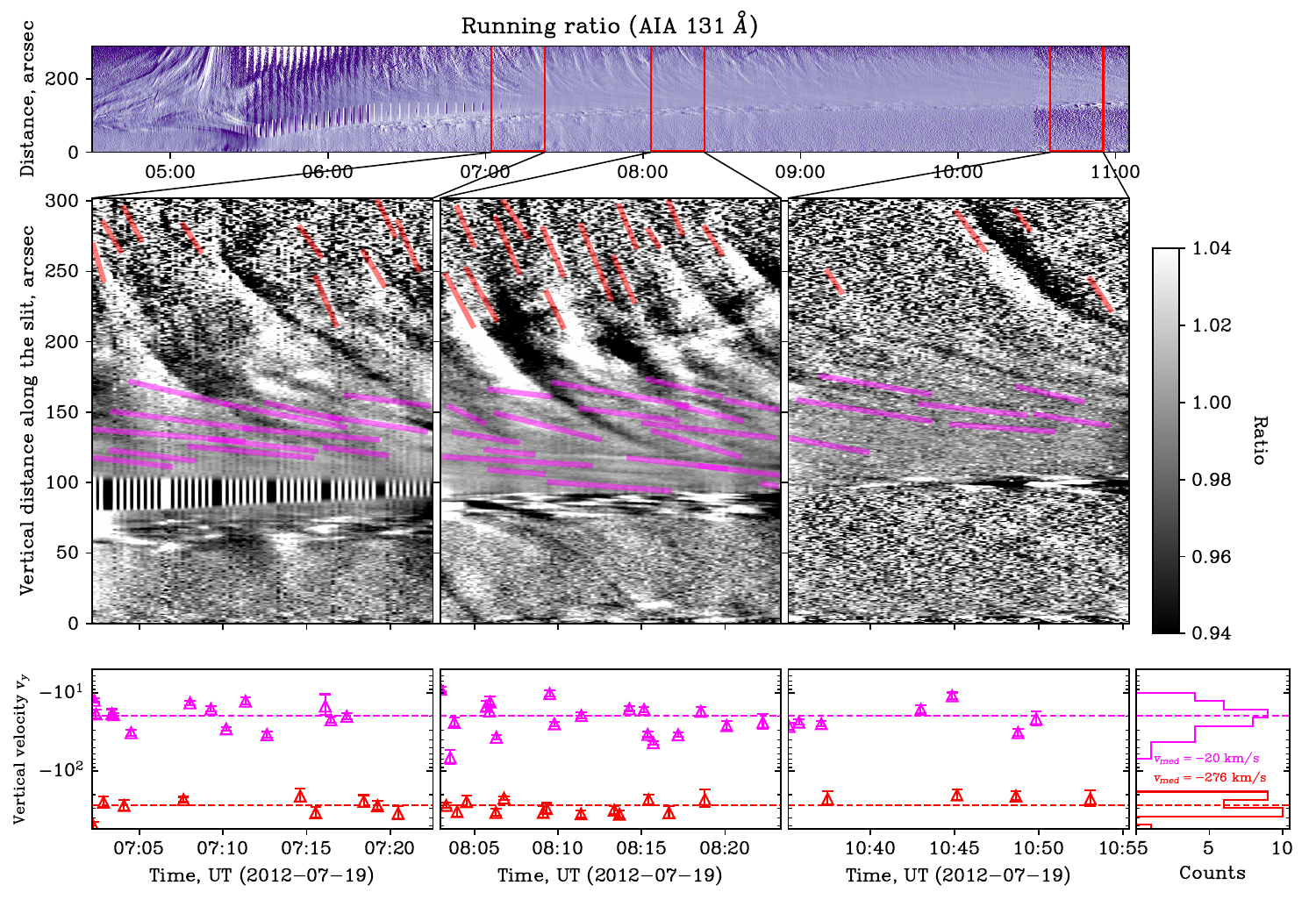}
\caption{The evolution of the plasma flows and flare loops in the 2012 July 19 flare. The upper panel covers the time period from the impulsive phase of the flare, including flux rope eruption at around 5:00 UT, and demonstrates a sustained reconnection process in the gradual phase and production of downward reconnection jets up to 11:00 UT. The three subpanels below are detrended {20 minute} windows, chosen to match the time coverage of the numerical model. The running ratio time--distance diagram in the right-most panel is extracted from the vertical slit introduced in figure \ref{fig:obs_slits}, while the central and left-most subpanels are extracted using the slit \# 2 defined in \citealt{liu_plasmoid_2013} to reach better alignment with the RCS position at earlier times. 
Linear fits of fast and slow plasma flows are shown in red and magenta triangles (lines), respectively. The lower panel shows velocity estimates for fast and slow downflows. The histogram on the right side shows the distribution of fast and slow downflows and median velocity estimates, shown on all panels as dashed horizontal lines.}  
\label{fig:obs_dwfls}
\end{figure*}

Reconnection theories and numerical experiments suggest that, due to the presence of complex structures and dynamics in the RCS, the plasma speeds in the RCS have a wide range of distribution, with only the fastest ones approaching the Alfv\'en speed in the inflow region \citep{liu_structure_2012, Haggerty2018, liu_ohms_2025}.
In order to identify velocity peaks at a given $y\text{-}z'$ slice corresponding to the reconnection downflows, we collapsed the velocity distribution along $z'$ axis and derived a one-dimensional profile of $v_y^{\rm max}(y)$ with height, where peaks are expected to correspond to the fastest downflow in the slice.
To simulate the finite thickness of the AIA 131\,\AA~ slit (about 5 pixels with a resolution of 0.6~\arcsec pix$^{-1}$, or $\sim$2 Mm in physical width), we extracted $N_{\rm slc} = 20$ slices that are contained within this slit thickness range. We then computed a $v_y^{\rm max}(y)$ profile for each $y$-$z'$ slice (shown in blue in Figure~\ref{fig:velmap}(b)), and determined an envelope curve for all slices (shown in red). To find correspondence between individual downflows and their time--distance response that will be further compared with synthetic EUV signatures, we stacked all $v_y^{\rm max}(y, t)$ profiles in a single time--distance diagram (right panel on Figure~\ref{fig:velmap}). This map allows one to identify the fastest reconnection downflows as they propagate from the X-line, seen as a dark region at $y\approx 0.9$, downward to the upper part of the above-the-loop-top region, which grows from $y\approx 0.3$ at the beginning to $y\approx 0.55$ at $t=7$.

Further, since the observed apparent plasma downflows involve all the reconnection outflows weighted by the density and temperature distributions along the line of sight, we also compute emission measure $\displaystyle \mathrm{EM}~=~\int n^2 dz'$ weighted velocity profiles $\langle v\rangle_{\rm EM}(y, t)$: 
\begin{equation}
\label{eq:emw_v}
\langle v\rangle_{\rm EM}(y, t) = \frac{\int n^2~T~v_y(y, z')~ H(j_z - j_z^{\text{min}}) dz'}{\int n^2~T~H(j_z - j_z^{\text{min}}) dz'},    
\end{equation}
\noindent
where $H$ is a Heaviside step function that filters out the grid cells with a current density lower than a threshold value $j_z^{\rm min}$. This is to ensure that the integration does not include background plasma outside the current sheet region, which usually has a lower temperature and does not contribute much to the observed EUV intensity in, e.g., the 94 and 131~\AA\ passbands. For each $y$-$z'$ slice within the 5$''$ thick slit (along $x'$), we compute an EM-weighted vertical velocity profile ${\langle v\rangle_{\rm EM}(y)}$, shown as light green curves in Figure~\ref{fig:velmap}(b). 
We then average the EM-weighted vertical velocity over all slices $\displaystyle {\langle v\rangle_{\rm EM}^{\text{slit}}(y, t)} = \frac{1}{N_{\rm slc}}\sum_i v_{\rm EM}^{( x'_i)}(y, t)$. The averaged height profile of $\langle v\rangle_{\rm EM}^{\text{slit}}$ is shown as a green curve in Figure~\ref{fig:velmap}(b). As can be seen from Figure~\ref{fig:velmap}(b), for all oblique 2D slices parallel to the slit plane, EM-weighted velocity does not exceed one-half of the maximum downflow velocity within the slit plane.

Figure \ref{fig:param_dist_los} illustrates density and temperature profiles within the oblique cut across the RCS, defined by slit position and line-of-sight direction. It can be seen that temperature and density distributions do not show similar behavior to the velocity profile. More notably, the highest velocity regions do not necessarily have a higher density or temperature. Therefore, plasma flows with different velocities can all contribute to the observed EUV brightness weighted by the density and temperature along the line of sight, rendering the EM-weighted velocity at nearly one-half of the maximum downflow velocities.

Comparisons of $\langle v\rangle_{\rm EM}$ with $v_{\rm max}$ shown in figures \ref{fig:param_dist_los} and \ref{fig:velmap}(b) indicate that without taking into account the plasma material distribution along the line of sight, the measurements of apparent velocity of plasma downflow may lead to a large underestimation of the reconnection outflow velocity, and Alfv\'en speed in the inflow region.

\subsection{Apparent Downflow Speeds in the Time--Distance Plots}\label{subsec:dwfltdist}\nopagebreak[4]
In this subsection, we further simulate synthetic EUV 131~\AA\ time series images using the time-dependent 3D MHD simulations, 
and then produce running ratio time--distance maps to best replicate the observational practice of tracking the plasma downflows.
The synthetic AIA 131~\AA\ time--distance map made at the same slit \#2 shown in Figure~\ref{fig:obs_slits} is displayed in Figure~\ref{fig:dwfls}(a). The running ratio time--distance map, in which the intensities shown at each time are produced by dividing the current image by the preceding one 12~s before, is shown in Figure~\ref{fig:dwfls}(b). 

We choose a 12 minute time interval that features multitudes of plasma downflows in the reconnection current sheet above the Y-points. We have identified 18 plasma downflows that manifest themselves as distinct bright tracks in the running ratio time--distance plot. 
The downflow tracks are identified by detecting local intensity peaks at a given time step within the height range of interest. To determine the speeds of the apparent downflows, we perform linear fits to derive the slopes of the tracks, which are translated into speeds. We note that there are also many tracks below the Y-points with less steep slopes (Figure~\ref{fig:dwfls}(b), magenta). They represent the contracting loops in the above-the-loop-top region with much slower speeds, with a median around $100$~km~s$^{-1}$. Knowing the exact location of the Y-points from the 3D MHD model helps us conclusively remove the possibility of tracking flows that are not associated with the reconnection outflows.

The time--distance derived speeds of plasma downflows (black triangles on Fig. \ref{fig:dwfls}({c}) are distributed in the range from 200~km~s$^{-1}$ to 600~km~s$^{-1}$, peaking approximately around 400~km~s$^{-1}$, as can be seen from the histogram shown in the figure~\ref{fig:dwfls}({d}). 
We then compare these results with the instantaneous velocities extracted directly from the MHD dataset.  By utilizing previously calculated maximum and EM-weighted downflow velocities, we derive stacked time--distance maps, such as one shown in Fig.~\ref{fig:velmap} (c) for maximum $v_y(y, t)$ along the slit. On that plot, the maximum velocity corresponds to the fastest downflow.


In a related manner, we compute a time--distance map for EM-weighted slit average ${\langle v\rangle_{EM}^{\text{slit}}}$ using equation~\ref{eq:emw_v}, for all downflows that cross the line of sight at a given height along the slit. 
From the histogram showing distribution of downflow speeds (Fig. \ref{fig:dwfls}({d})), we could see that the EM-weighted speeds ${\langle v\rangle_{EM}^{\text{slit}}}$ generally agree with the time--distance-map-derived speeds, demonstrating that the brightness features in the time--distance analysis are EM weighted. 


\subsection{Comparison with the observed downflows}

To validate the plasma downflow speeds derived using the synthetic time--distance plots based on the numerical model, we applied an identical fitting approach to measure the downflow speeds in the observations of the 2012 July 19 solar flare, which we have used to tune the model parameters. We selected three 20~minute long time windows in the gradual phase of the event that host multitudes of detectable plasma downflows. The last time interval (10:35--10:55~UT) also overlaps with the time interval used for analysis of imaging observations in Section \ref{sec:morphology}. Earlier windows at 7:00~UT and 8:00~UT feature more downflows, which are also faster, possibly indicating more intense and dynamic reconnection processes earlier in the flare energy release. 

To estimate velocities of the downflows from SDO/AIA observations, we adapted an approach from \citet{yuMagneticReconnectionPostimpulsive2020}, which uses the running ratio of consecutive frames that is detrended over a period exceeding the characteristic time of the downflow appearance, in order to improve the contrast of the time--distance diagram. We also used an expanding slit with an opening angle $\alpha\simeq3^{\circ}$ to increase the signal-to-noise ratio at higher altitudes.
Results of the downflow fits are shown in Figure~\ref{fig:obs_dwfls}. The downflow speeds generally fall into the range of 200--400~km~s$^{-1}$, consistent with those derived from the synthetic AIA 131~\AA\ images. 
In addition to that, we determined median velocities of contracting loops, which appear to be of the order of magnitude less than the faster plasma flows located higher in the corona.

It should be noted that the model-derived time--distance plot clearly demonstrates more downflow tracks, as compared to the observational ones. The more abundant downflows {are} because the synthetic EUV images are computed directly from the simulation output, which have almost infinite dynamic range to show a much larger number of varying features in the running ratio images. 

\vspace{1em}
\section{Discussion and Conclusion}\label{sec:conclusion}The above-the-loop-top region at the bottom of the RCS, where the macroscopic Y-point is located, hosts numerous dynamic structures, including fast-mode termination shocks and a variety of plasma flows and oscillations \citep{chenParticleAccelerationSolar2015, takasaoAbovethelooptopOscillationQuasiperiodic2016, shenDynamicalBehaviorReconnectiondriven2018, shenOriginUnderdensePlasma2022}. In particular, its ``magnetic bottle'' configuration \citep{chenMeasurementMagneticField2020, chenEnergeticElectronsAccelerated2024} may help facilitate its key role in particle acceleration and confinement 
\citep{kruckerMeasurementsCoronalAcceleration2010,kruckerPARTICLEDENSITIESACCELERATION2013b, kongAccelerationConfinementEnergetic2019, arnoldElectronAccelerationMacroscale2021b, liModelingElectronAcceleration2022, fleishmanSolarFlareAccelerates2022, liEnergyConversionElectron2025}, 
and plasma heating \citep{reevesHotPlasmaFlows2020, gouThermalPropertiesCurrent2024a} in flares. This location also features plasma downflows, whose speeds have been used to infer the rate of magnetic reconnection. As such, the Y-points serve as an important anchor to relate the energy release and particle acceleration processes and plasma dynamics to the underlying magnetic geometry. Reliable observational diagnostics of the Y-point locations and a correct interpretation of the associated plasma outflows are thus highly desirable.

In this work, we have examined how the observed candle-flame-shaped structure and apparent plasma downflows as measured in the time--distance plots can be used to inform the location of the reconnection Y-points and reconnection outflow speeds, by utilizing the state-of-the-art numerical 3D MHD simulation developed by \citet{shenOriginUnderdensePlasma2022}. Comparisons to actual observations are made using EUV and SXR time series imaging observations of the candle-flame-shaped eruptive flare on 2012 July 19 obtained by SDO/AIA and Hinode/XRT. Our main findings include the following:
\begin{enumerate}
    \item By directly examining the current density and magnetic field in the 3D numerical model, we pinpoint the location of the Y-points derived from both current sheet bifurcation and local magnetic field minimum.
    We then produce realistic synthetic EUV and SXR images by adjusting the model to match the flare parameters and viewing geometry of the 2012 July 19 flare, which successfully reproduces the observed candle-flame-shaped flare morphology. 
    We find that most of the projected Y-points are, in fact, distributed well below the apparent tip of the cusp feature. This location corresponds to the region where the current sheet in the model bifurcates and broadens, and plasma density rises at a sharper exponential rate.
    
    \item The RCS hosts flows with a broad distribution of speeds, with the fastest ones reaching about 80\% of the Alfv\'en velocity in the inflow region $v_A^{\rm in}$. However, in agreement with the observations, the speeds of apparent plasma downflows above the Y-points, as derived from the synthetic AIA 131~\AA\ time--distance plots, are only $\sim\!0.5v_A^{\rm in}$. For downflows below the Y-points, they are much slower, at only $\sim\!0.1v_A^{\rm in}$ (see Fig.~\ref{fig:velmap}(b)).
    
\end{enumerate}

Our results show that the apparent tip of the cusp is not a reliable means for determining the location of the macroscopic Y-points. In fact, in our experiment, a large portion of the spike feature at the very tip corresponds to the bottom part of the RCS, and the actual Y-points are ``buried'' well below the visible tip. 
A better and more qualitative means for identifying the general location of the macroscopic Y-points based on EUV/SXR imaging is to make a brightness profile along the direction of the RCS. From the profile, we can then use the transition region from an exponential brightness drop at lower heights to a slow-varying ``plateau'' at higher altitudes as the proxy for the general Y-point location. However, this method can only be used for flare events with its current sheet viewed nearly edge-on. Also, any nonuniformity in 3D may play a role in altering the brightness distribution. 


Alternatively, to determine the location of the bottom of the RCS, time-resolved measurements of plane-of-sky motions could be used. 
Time--distance diagrams provide a clear visualization of the transition from fast plasma downflows to slow contracting motion of postflare loops. This apparent break in plasma flow speed identifies the topological transition from quasi-parallel magnetic field geometry to closed, and traces the height and temporal development of the Y-line. This method works well for flares close to the limb and appears less sensitive to the current sheet orientation (edge-on/face-on). 

Another key aspect of our study is to investigate whether the apparent plasma downflow speeds, as derived from the time--distance plots of the synthetic EUV time series images, can be used as a reliable proxy for the Alfv\'{e}n speeds in the reconnection inflow region $v_A^{\rm in}$. Although previous observational studies have alluded that they may be slower than the presumed Alfv\'{e}n speeds by {up to} an order of magnitude \citep{savageQUANTITATIVEEXAMINATIONLARGE2011, longcopeEvidenceDownflowsNarrow2018a}, our results, which are based on ``ground truth" parameters derived from the 3D MHD model, not only provide a solid confirmation, but also offers physical insights into the origin of such a discrepancy.

First and foremost, where we track the supra-arcade plasma downflows matters. If the tracked plasma downflows are located below the Y-points and do not reside in the RCS, the plasma motions are much slower due to the greater cross-sectional area as they exit the narrow current sheet and the strong resistance they experience from the underlying flare loops. Our simulation results show that these flows are only $\sim\!0.1v_A^{\rm in}$. As a result, the inflow Alfv\'{e}n speed can be underestimated by an order of magnitude just because the ``wrong'' plasma outflows are tracked.  

Second, the flows in the RCS region are modulated by various structures and complex dynamics associated with, e.g., non-laminar flows, flows originating at different heights due to the meandering X-line in 3D, and also the plasmoid formation that interrupts the outflow structure and leads to the fragmentation of the outflows. Analogous behavior of reconnection outflows was observed in MHD and particle-in-cell simulations of general reconnection geometries \citep{huangTurbulentMagnetohydrodynamicReconnection2016, guoMagneticEnergyRelease2021}. Our 3D MHD simulations in an eruptive flare context show that these flow speeds show a large range of variation, with a maximum speed of $\sim\!0.8v_A^{\rm in}$, but the slowest ones approach zero.  
The apparent plasma downflow speeds measured in the time--distance plots are contributed by all such flows in the line of sight, weighted by their density and temperature. As a result, the line-of-sight integration renders the measured speeds in our experiment to be substantially lower than the highest achievable speed, at only around 0.5$v_A^{\rm in}$. We also need to emphasize that, due to the sensitivity of the observed EUV brightness to the detailed thermodynamic processes, this figure may vary on a case-by-case basis. 




Last but not least, we note that projection effects can further reduce the apparent speeds of the plasma downflows by a factor of $\sin\alpha$ (where $\alpha$ is the angle between the line of sight and the direction of the downflows $y$). 
In the case of the 2012 July 19 event, with an estimated small tilt angle toward the observer, the projection effect is negligible. However, this effect can be more profound for events with a much greater tilt.


For reasons discussed above, in sum, the practice of using apparent plasma downward speeds $v_{\rm d}$ as a proxy of Alfv\'en speed can easily lead to a severe underestimation of up to an order of magnitude, which, in turn, gives an artificially high dimensionless reconnection rate. As such, special care must be taken when performing this type of analysis. Also, our results suggest that the reconnection rate reported in the literature using this method may be overestimated by at least a factor of 2 in the best-case scenario, to more than an order of magnitude if they used the ``wrong'' type of plasma downflows as the proxy. It may explain why sometimes the reported reconnection rate exceeds the theoretically achievable value of $M_A\approx 0.1$. 
\begin{acknowledgments}
I.O. carried out the majority of the work during his participation in the Smithsonian Astrophysical Observatory (SAO) Predoctoral Program. I.O. would like to thank Drs. Yuhao~Chen and~Jun~Lin for useful discussions and comments. I.O., S.F., B.C., and S.Y. acknowledge support by NSF grants AST-2108853 and AGS-2334931 to the New Jersey Institute of Technology (NJIT), NASA grant 80NSSC20K1318 to NJIT, and NASA grant NNH240B72A through subcontract C5509 from DOE/LANL. In addition, S.Y. acknowledges support by NASA grant 80NSSC24K1242 to NJIT. C.S. acknowledges the support from NASA through grant Nos. 80NSSC21K2044 and 80NSSC25K7707. X.L. and F.G. acknowledge support from NASA through grant NNH240B72A and from the National Science Foundation through grants AST-2107745 and AGS-2334930.
\end{acknowledgments}

%

\vspace{5mm}
\facilities{SDO (AIA), STEREO (EUVI), Hinode (XRT)}


\software{sunpy, 
            scikit-image, SolarSoft
          }

\newpage
\appendix

\section{Identification of Y-points}
\label{app:yptsrec}

We extracted 90 $x\text{-} y$ plane slices of $j_z$ along the $z$-axis and computed the topological structure of the current density distribution to pinpoint the bifurcation site. The binary image representing the geometry of the current sheet cross section was obtained after applying the skeletonization algorithm \citep{zhangFastParallelAlgorithm1984} available in the \texttt{scikit-image} image processing library \citep{walt_scikit-image_2014}\footnote{{\url{https://scikit-image.org/docs/stable/}}}. For the skeletonized image of the current density, we identified branching points and picked the uppermost one that has no branches intersecting a horizontal line at the same $y$ coordinate. The lines with unconnected ends within the ALT region were excluded from the binary images to avoid recognition artifacts. To avoid interferences from the fine structures within the RCS (such as plasmoids and geometry disturbances) detectable at the full resolution of the initial dataset, the slices were extracted from current density distributions determined from the simulation output and further image resolution reduction with a factor of $n=3$.

The profiles of the simulated magnetic field extracted along the current sheet typically show a flat minimum region that could host multiple local minima. To extract 1D profiles of the magnetic field along the vertical direction, we used a Gaussian profile to fit the thickness of the RCS in the vicinity of the RCS bifurcation point and determine its central coordinate for a particular simulation timeframe. To account for the finite thickness of the RCS, we extracted several vertical slices and found the mean value of the magnetic field along the $x$-coordinate, and then identified the local magnetic field minimum along the vertical profile of $B(y)$.
To associate the local magnetic field minimum with a reconnection Y-point, we picked the detections of the minima approaching the bifurcation site from the lower side.

\clearpage 
\section{Y-POINT EVOLUTION AND FINE PLASMA STRUCTURES INSIDE the RCS}
\begin{figure}[H]
\begin{interactive}{animation}{ypt_evolution_and_cs_structure.mp4}
\includegraphics[width=0.65\textwidth]{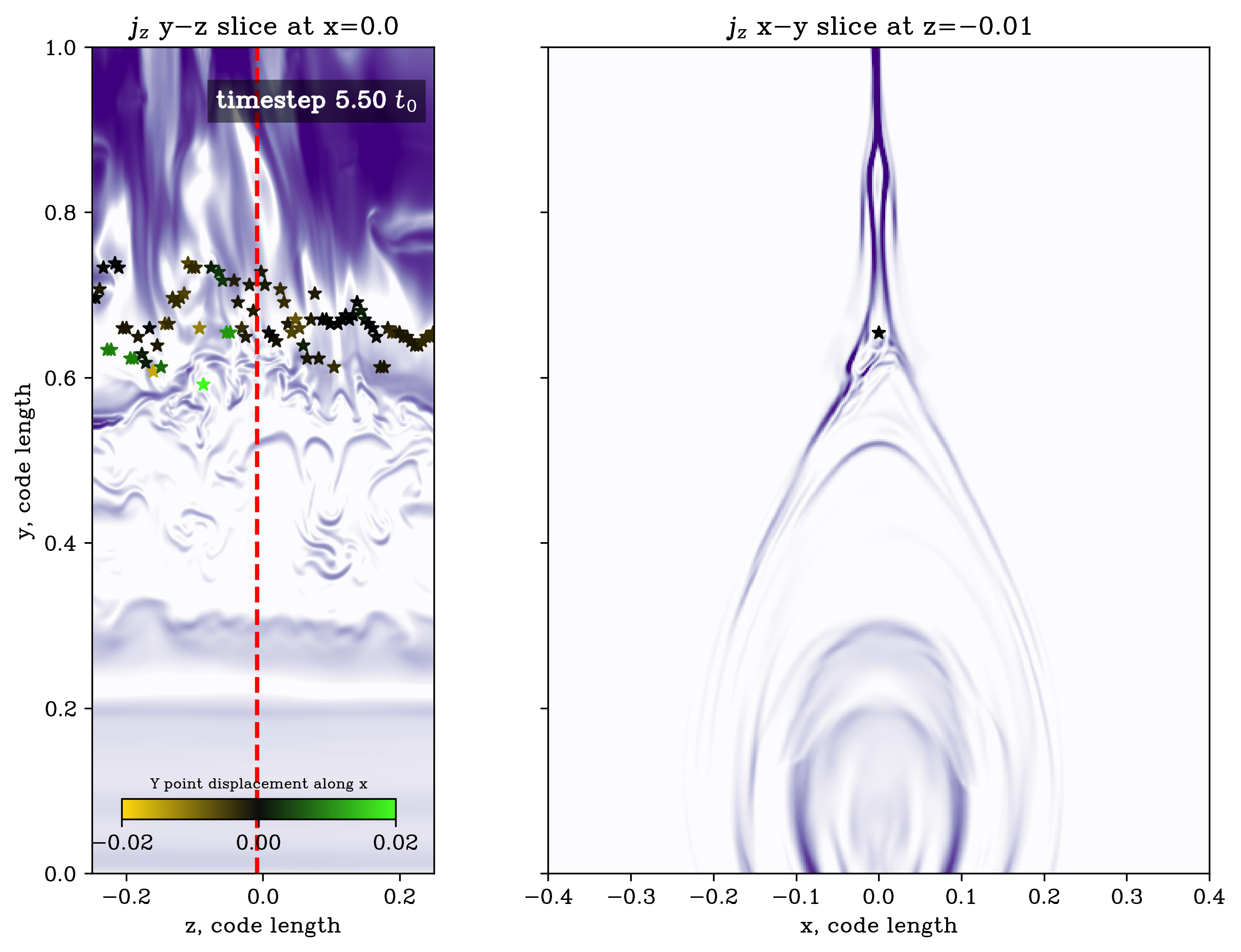}

\end{interactive}
\caption{{This animation (available in the HTML version) shows the evolution of the Y-line (in the $y-z$ plane) and development of perturbations inside the RCS. The left panel shows the $y-z$ slice of current density extracted along $x=0$ throughout the MHD simulation at the time step shown in the panel legend. Y-points are overplotted over the current density map (purple). Color code displays Y-point displacement along $x$. The right panel shows the corresponding slice in the $x-y$ plane. The red dashed vertical line in the left panel demonstrates the $z$-coordinate at which the corresponding $x-y$ slice is extracted and plotted in the right panel.}}
\label{fig:ypt_animation}
\end{figure}

Here we consider the temporal evolution of Y-points and the formation of magnetic perturbations and plasmoids in the developing RCS (see Fig. \ref{fig:ypt_animation}).
\label{app:ypt_rcs_scan}

\vspace*{\fill} 
\clearpage 

\section{Alignment of MHD model to 2012 July 17 event geometry}
\label{app:align}

In order to perform accurate model-to-data comparisons, the MHD model was projected into the viewing geometry, accounting for observed arcade location on the solar surface and orientation towards the observer (SDO/Hinode observatories).
Additionally, the three-dimensional MHD model utilized in this study is defined with a coordinate system as follows:
\begin{enumerate}
    \item[] $x$ axis--parallel to the loop plane, connecting loop footpoints
    \item[] $y$ axis--parallel to the loop plane, running up through the flare cusp
    \item[] $z$ axis--perpendicular to the loop plane, running through the flare arcade
\end{enumerate}

\begin{figure*}[ht!]
\centering
\includegraphics[width=0.95\textwidth]{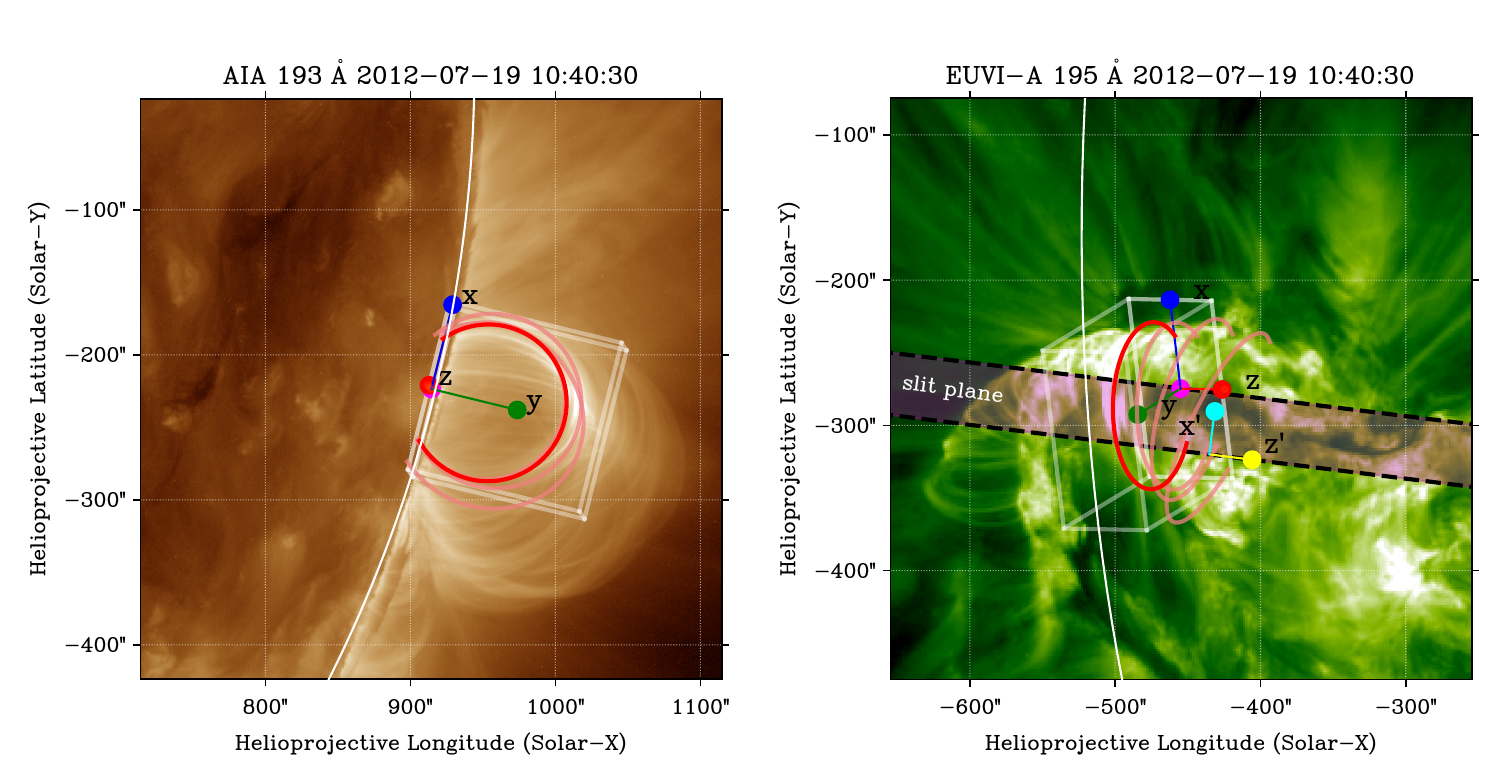}  
\caption{Aligning the MHD simulation dataset with the observations of postflare loop configuration using joint SDO and STEREO EUV data. The red line indicates the geometry of the postflare loop derived from the fit {and used to determine the line-of-sight direction for synthetic imaging}. {Loops shown in pink show fits of adjacent EUV loops used to estimate the range of angles that the line of sight makes with the flare arcade}. The colored line segments (blue, green, red) indicate the MHD coordinate system principal axes ($x$, $y$, $z$), and the white box indicates the orientation of the MHD simulation volume. The magenta-colored filling shows the slit plane as seen from the STEREO perspective, which is bordered by AIA lines of sight that define a vertical slice along the flare current sheet. The slit is aligned with the vertical axis of the MHD dataset $y'\equiv y$. The $x'$ axis is perpendicular to the $y$-$z'$ plane.} 
\label{fig:alignment}
\end{figure*}

To determine components of the line-of-sight vector, we first define the location of the flare arcade in a Heliographic Stonyhurst coordinate frame.
An array of coordinates in the Stonyhurst heliographic frame \citep{thompson_coordinate_2006} representing points along a single flare loop was aligned to visually match the observable flare loop geometry in two perspectives of the 2012 long-duration event: one from the SDO/AIA instrument and the other from the STEREO/EUVI instrument (see fig. \ref{fig:alignment}).
These coordinates are generated using the software module \texttt{Coronal Loop Builder}\footnote{\url{https://github.com/sageyu123/CoronalLoopBuilder}} (CLB).

Once these loop coordinates were secured and converted to the Heliocentric-Cartesian frame using the \texttt{sunpy} package, the principal axes of the loop plane could be derived. The coordinates for both footpoints of the CLB loop become vectors pointing out from the solar center. A third vector is introduced, pointing from the solar center to the top of the CLB loop. 
The cross product of these two in-plane vectors is chosen to align with the $z$-axis of the MHD coordinate frame. Next, the vector connecting loop footpoints is defined along the $x$-axis of the simulation domain, and the $y$-axis is chosen along the vertical direction defined by the cross product between the $z$ and $x$-axes. 

With the principal axes of the MHD frame resolved, a transformation matrix was constructed so that the Heliocentric-Cartesian coordinates for the observer of the projection could be easily converted into the MHD frame. Observer's coordinates are defined from the header of the FITS file used for model-to-data comparison.  The observer is defined to be located in the negative direction of the $z$-axis of the Heliocentric Cartesian coordinate system (similarly, negative $\zeta$ in Helioprojective Cartesian). Taking the product of the observer's location vector with the transformation matrix provides the line-of-sight vector expressed in MHD simulation domain coordinates. 

With the line-of-sight vector defined in the MHD coordinate frame, we compute a synthetic image projection with the help of the 3D data visualization package \texttt{yt} \citep{turk_yt_2011}.
The final step of creating a properly aligned synthetic map was to scale and translate the synthetic data to the correct location in the final  Helioprojective-Cartesian frame. This was accomplished by projecting the origin of the MHD box into the Helioprojective-Cartesian frame and displacing the image by the difference in pixels between the MHD origin and the CLB loop coordinates, also projected into the Helioprojective-Cartesian frame. Scale factor was determined using the ratio between the physical distance between flare footpoints and the related distance between footpoints in the MHD model (in Mm) after conversion, assuming characteristic length scale $L_{\text{char}} = 150$~Mm. Derived physical units, including the timescale, were scaled accordingly, preserving the Alfv\'en speed.  

\bibliography{oparin, sample631}{}
\bibliographystyle{aasjournal}



\end{document}